\documentclass[preprint,aps]{revtex4}
\begin{document}
\title{SUSY QM from Three Domain Walls in a Scalar Potential}
\author{R. de Lima Rodrigues$^{(a)},$
A. F. de Lima$^{(b)},$\\
 E. R. Bezerra de Mello$^{(c)},$
and V. B. Bezera$^{(d)}$\\
{}$^{(a)}$Centro Brasileiro de Pesquisas F\'\i sicas\\
Rua Dr. Xavier Sigaud, 150\\
CEP 22290-180, Rio de Janeiro-RJ, Brazil\\
 {}$^{(b)}$ Departamento de F\'\i sica, Universidade Federal de
Campina Grande \\
 {}$^{(c,d)}$ Departamento de
F\'\i sica, Universidade Federal da
Para\'{\i}ba \\
58.051-970 - Jo\~ao Pessoa - PB}

\begin{abstract}
We investigate the linear classical stability of
Bogomol'nyi-Prasad-Sommerfield (BPS)  on three domain wall
solutions in a system  of three coupled real scalar fields, for a
general positive potential with a square form. From a field
theoretic superpotential evaluated on the domain states, the
connection between the supersymmetric quantum mechanics involving
three-component eigenfunctions and the stability equation
associated with three classical configurations is elaborated.

\vspace{2cm}


\vspace{4cm}
 \small{To appear in proceedings of the Fifth
International Conference on Mathematical Methods in
Physics--IC2006. April 24-28 2006.  Centro Brasileiro de Pesquisas
F\'\i sicas, Rio de Janeiro, Brazil}

\end{abstract}

\maketitle

\section{Introduction}

The kink solution of a field theory is an example of a soliton in
1+1 dimensions \cite{Jac,Raja,Cole,Bala,Weinberg}.
From field theoretic superpotential \cite{R95} evaluated on the
domain states, the algebraic framework of supersymmetry in quantum
mechanics (SUSY QM), as formulated by Witten \cite{W}, may be
elaborated. The SUSY QM generalization of the harmonic oscillator
raising and lowering operators has several applications
\cite{Fred}. The generalization of SUSY QM for the case of matrix
superpotential, is well known in the literature for a long time.
See, for example, for one-dimensional systems the works in Ref.
\cite{rafa01} about non-relativistic quantum systems, in
\cite{vvr02,vachas04} about one single field  and in
\cite{RPV98,guila02,ger06} with two-field and in
\cite{RPV95} on a three-field potential models in 1+1
dimensions.

The classical configurations with domain wall solutions  are
bidimensional structures in 3+1 dimensions. They are static,
non-singular, classically stable Bogomol'nyi \cite{Bogo} and
Prasad-Sommerfield \cite{PS} (BPS) soliton (defect)
configurations, with finite localized energy associated with a
real scalar field potential model. Domain walls have several
applications in condensed matter and cosmology. The BPS states are
classical configurations that satisfy the first order differential
equations and the second order differential equations (equations
of motion).

Recently, marginal stability and the metamorphosis of BPS states
have been investigated \cite{Shif01b}, via SUSY QM, with a
detailed analysis for a 2-dimensional $N=2-$Wess-Zumino model in
terms of two chiral superfields, and composite dyons in
$N=2$-supersymmetric gauge theories. Domain walls have been
recently exploited in a context that stresses their connection
with BPS-bound states. While Rajaraman has applied the trial orbit
method for the equation of motion, here we use the trial orbit
method for the first order differential equations associated to
three real scalar fields. However, for solitons of three coupled
scalar fields there are no general rules for finding analytic
solutions since the nonlinearity in the potential increases the
difficulties to obtain the solutions of the BPS equations and field equations.

This paper is organized as follows: In Section II, we investigate
domain walls configurations for three coupled scalar fields, and
supersymmetric non-relativistic quantum mechanics with
three-component wave functions is implemented. In Section III, a
scalar potential model of three coupled scalar fields is
investigated. Our Conclusions are presented in Section IV.

\section{Three coupled scalar fields}

\paragraph*{}

We consider the classical soliton solutions of three coupled real
scalar fields in 1+1 dimensions. They are static, nonsingular,
classically stable and finite localized energy solutions of the
field equations.

Recently, a superfield formulation of the central charge anomaly
in quantum corrections to soliton solutions with N=1 SUSY has been
investigated \cite{KS04}. The superaction in terms of the
superfield and the superpotential in superspace
$(x^{\mu};\Theta_{\alpha}), \alpha=1,2,$ is written as

\begin{equation}
\label{SA} S_{N=1}=\frac 12\int
dx^2d^2\Theta\left(\sum_{i=1}^3\bar D\Phi_i D\Phi_i + W(\Phi_1,
\Phi_2, \Phi_3)\right),
\end{equation}
where $D$ is the supercovariant derivative given by

\begin{equation}
\label{SD} D=\partial_{\bar\Theta}+i\bar
\Theta\gamma^{\mu}\partial_{\mu}
\end{equation}
and the $\gamma^{\mu}$ are represented in terms of the Pauli
matrices, $\gamma^0=\sigma_y$ and $\gamma^1=i\sigma_x.$ The
superpotential $W(\Phi_1, \Phi_2, \Phi_3)$, yields the
component-field potential $V(\phi_i)$, where $\Phi_i$ are chiral
superfields which, in terms of bosonic ($\phi_i$), fermionic
$(\psi_i)$ and auxiliary fields $(F_i)$, are $\Theta_i-$expanded
as shown below:

\begin{equation}
\label{SF}
  \Phi_i =\phi_i+\bar\Theta_i\psi_i+\frac{\Theta_i\bar\Theta_i}{2}F_i,
\end{equation}
where $\Theta_i$ and $\bar{\Theta_i}=\Theta_i^*$ are Grassmannian
variables.

The Lagrangian density for such nonlinear system in the natural
system of units $(c=\hbar=1)$, in (1+1)dimensional space-time with
Lorentz invariance, in terms of the bosonic fields only, is
written as

\begin{equation}
{\cal L}\left(\phi_j, \partial_{\mu}\phi_j\right) =
\frac{1}{2}\sum^3_{j=1}\left(\partial_{\mu}\phi_j\right)^2
-V(\phi_j),
\end{equation}
where $\partial_{\mu}=\frac{\partial}{\partial z^{\mu}}, \quad
z^{\mu}=(t,z), \quad \phi_j =\phi_j(t,z), \quad
(j=1,2,3) $ are real scalar fields and $\eta^{\mu\nu}=(+,-)$ is the
metric tensor. The potential
$V(\phi_j)=V(\phi_1,\phi_2,\phi_3)$ is any positive definite
functional of $\phi_j$. The general classical configurations obey the
equation bellow:

\begin{equation}
\frac{\partial^2}{\partial t^2}\phi_j - \frac{\partial^2}{\partial
z^2}\phi_j +\frac{\partial }{\partial \phi_j}V=0,
\end{equation}
which, for static soliton solutions, is equivalent to the following
system of nonlinear second order differential equations

\begin{equation}
\label{EMG} \phi_j^{\prime\prime}=\frac{\partial }{\partial
\phi_j}V, \quad (j=1,2,3),
\end{equation}
where prime represents differentiation with respect to space
variable. There is, in literature, a trial orbit method for finding
static solutions for certain positive potentials, which constitutes
a "trial and error" technique \cite{Raja}.

We can get the masses of the bosonic particles, using the results
above, from the second derivatives of the potential:

\begin{equation}
\label{massa}
  m^2_{\phi_i}\equiv
\frac{\partial^2V}{\partial\phi_i^2}\left|_{z\rightarrow\pm\infty}\right.
, \quad i=1,,2,3.
\end{equation}

\subsection{Linear Stability and SUSY QM}

\paragraph*{}

Since the potential $V(\phi_j)$ is positive it can be written in the
following square form, analogous to the case with only a single field
\cite{R95},

\begin{equation}
\label{EP}
  V(\phi_j)=
V(\phi_1,\phi_2,\phi_3)=\frac{1}{2}\sum^3_{j=1}U_j^2(\phi_1,\phi_2,\phi_3),
\quad U_j(\phi_1,\phi_2,\phi_3) \equiv\frac{\partial
W}{\partial\phi_j}, \quad (j=1,2,3).
\end{equation}
In this case, one can write the total energy

\begin{equation}
\label{ET3} E=\int_{-\infty}^{+\infty} dz\frac
12\left[\left(\phi_1^{\prime}\right)^2
+\left(\phi_2^{\prime}\right)^2 +\left(\phi_3^{\prime}\right)^2 +
2V(\phi_i)\right],
\end{equation}
in the  BPS form, consisting of a sum of squares and surface terms,

\begin{equation}
\label{ETG} E=\int_{-\infty}^{+\infty} dz \left(\frac
12(\phi_1^{\prime}-U_1)^2+\frac 12(\phi_1^{\prime}-U_2)^2+ \frac 12
(\phi_3^{\prime}-U_3)^2+\frac{\partial}{\partial z}W \right)
\end{equation}
which are always positive. Thus,
the lower bound of the energy (or classical mass) is given by the
fourth term, viz.,

\begin{equation}
\label{Ebogo}
E\geq\left|\int_{-\infty}^{+\infty} dz
\frac{\partial}{\partial z}W[\phi_1(z), \phi_2(z),
\phi_3(z)]\right|,
\end{equation}
where the superpotential $W=W[\phi_1(z), \phi_2(z), \phi_3(z)]$ will be
discussed below. The BPS mass bound of the
energy resulting in a topological charge is given by

\begin{equation}
\label{EBPS} E_{BPS}=T_{ij}=|W[M_j]-W[M_i]|,
\end{equation}
where $M_i$ and $M_j$ represent the BPS vacuum states and are the
extrema of $W.$.

In this case the BPS states satisfy the following set of first order
differential equations associated to three real scalar fields:

\begin{equation}
\label{CBPS}
  \phi_j^{\prime}=U_j(\phi_1,\phi_2,\phi_3).
\end{equation}
Now, let us analyze the classical stability of the soliton solutions
in this nonlinear system, which can be investigated by considering
small perturbations around $\phi_1 (z), \phi_2 (z)$ and $\phi_3
(z)$:

\begin{equation}
  \phi_j(t,z)=\phi_j(z)+\eta_j(t,z), \quad (j=1,2,3),
\end{equation}
where the fluctuations $\eta_j(t,z)$ can be expanded in terms of the normal
modes as

\begin{equation}
\eta_j (t,z) = \sum_n \epsilon_{j,n} \eta_{j,n} (z) e^{i\omega_{j,n}
t}, \quad \omega_{1,n}=\omega_{2,n}=\omega_{3,n}=\omega_n,
\end{equation}
with $\epsilon_{j,n}$ being real constant coefficients. Thus, the
stability equation for the field becomes a Schr\"odinger-like
equation of a three-component eigenfunction $\Psi_{n}$,

\begin{equation}
\label{SE} {\cal H} \Psi_{n} = \omega_{n}^2\Psi_{n}, \quad n=0, 1,
2, \cdots,
\end{equation}
where

\begin{equation}
\label{EFH1} {\cal H}\equiv-{\bf I}\frac{d^2}{dz^2}+V_F(z),
\end{equation}
with {\bf I} being the 3x3 identity matrix. The term $V_F(z)$ is the 3x3
fluctuation Hessian matrix and the excited mode is given by

\begin{equation}
\label{EFH} \Psi_{n}(z)=\left(
\begin{array}{ccc}
\eta_{1,n}(z) \\
\eta_{2,n}(z) \\
\eta_{3,n}(z)
\end{array}\right).
\end{equation}
Since

\begin{equation}
V_{Fij}=\frac{\partial^2}{\partial \phi_i\partial\phi_j}V
=\frac{\partial^2}{\partial \phi_j\partial\phi_i}V=V_{Fji}
\end{equation}
we see that  ${\cal H}$ is Hermitian. Hence the eigenvalues
$\omega^2_n$ of ${\cal H}$ are real.

The Schr\"odinger-like equation (\ref{SE}) and the Hessian matrix
$V_F(z)$ are obtained by considering a Taylor expansion of the
potential $V(\phi_j)$ in terms of $\eta_j(t,z)$ and retaining the
first order terms in the equations of motion.

From (\ref{EFH1}) we find a bilinear form of ${\cal H}$ given by

\begin{equation}
{\cal H}= {\cal A}^+ {\cal A}^-,
\end{equation}
where

\begin{equation}
{\cal A}^{\pm}=\pm\hbox{{\bf I}} \frac{d}{dz}+\hbox{{\bf W}}(z),
\quad {\cal A}^+=({\cal A}^-)^{\dagger}, \quad \hbox{{\bf
W}}^{\dagger}(z)=\hbox{{\bf W}}(z).
\end{equation}

Using these first order differential operators that appear in
analysis of classical stability associated to a single field
\cite{R95}, we find

\begin{equation}
\left({\cal A}^+ {\cal A}^-\right)_{jj}
=-\frac{d^2}{dz^2}+\frac{\partial^2}{\partial\phi^2_j}V,
\end{equation}
which are exactly the diagonal element of ${\cal H}$. Therefore,
it is easy to show that the linear stability is satisfied, which means that

\begin{equation}
\omega^2_n = <{\cal H}>=<{\cal A}^+{\cal A}^->= ({\cal
A}^-{\tilde\Psi}_{n})^{\dag}({\cal A}^-{\tilde\Psi}_{n})\geq 0,
\end{equation}
as was stated. Also, the ground state of ${\cal H}$ becomes

\begin{equation}
\label{ZMG}
\Psi_-^{(0)}(z)=\left(
\begin{array}{ccc}
U_1[\phi(z)] \\
U_2[\phi(z)] \\
U_3[\phi(z)]
\end{array}\right)
\end{equation}
which represents the bosonic three-component zero mode.

The 3x3-matrix superpotential satisfies the following Ricatti
equation associated to the non-diagonal fluctuation Hessian
$V_{F}(z)$:

\begin{equation}
\label{ER}
\hbox{{\bf W}}^2+\hbox{{\bf
W}}^{\prime}=V_{F}(z)=\left(
\begin{array}{ccc}
V_{F11}(z) &  V_{F12}(z)&  V_{F13}(z)\\
V_{F12}(z) & V_{F22}(z) &  V_{F23}(z)\\
V_{F13}(z) & V_{F23}(z) &  V_{F33}(z)
\end{array}\right)_{\vert\phi_i=\phi_i(z)}.
\end{equation}

The Ricatti equation (\ref{ER}) holds the BPS states only.
According to the Witten' SUSY model \cite{W,Fred}, we have

\begin{equation}
\label{A+-}
\Psi_{\hbox{SUSY}}^{(n)}(z)=\left(
\begin{array}{cc}
\Psi_-^{(n)}(z) \\
\Psi_+^{(n)}(z)
\end{array}\right)_{1\hbox{x}6},
\end{equation}
where  $\Psi_{\pm}^{(n)}(z)$ are three-component eigenfuctions. In
this case, the graded Lie algebra of the supersymmetry in quantum
mechanics for the BPS states may be readily realized as

\begin{equation}
\label{ofs}
 H_{SUSY} = [Q_- ,Q_+ ]_+ = \left(
\begin{array}{cc}
{\cal A}^+ {\cal A}^- & 0 \\
0 & {\cal A}^- {\cal A}^+
\end{array}\right)_{6\hbox{x}6}= \left(
\begin{array}{cc}
{\cal H}_- & 0 \\
0 & {\cal H}_+
\end{array}\right),
\end{equation}

\begin{equation}
\label{41} \left[H_{SUSY} , Q_{\pm}\right]_- = 0 = (Q_-)^2 =
(Q_+)^2,
\end{equation}
where $Q_{\pm}$ are the 6x6 supercharges of Witten SUSY model,
viz.

\begin{equation}
\label{SC} Q_- = \sigma_-\otimes{\cal A}^-, \quad
Q_+=Q_-^{\dagger}= \left(
\begin{array}{cc}
0 & {\cal A}^+ \\
0 & 0
\end{array}\right)=\sigma_+\otimes{\cal A}^+,
\end{equation}
with the intertwining operators, ${\cal A}^{\pm}$, in terms of
3x3-matrix superpotential, are given by Eq. (\ref{A+-}) and
$\sigma_{\pm}=\frac 12(\sigma_1\pm i\sigma_2),$ where $\sigma_1$
and $\sigma_2$ are the Pauli matrices. Of course, the bosonic sector
of $H_{SUSY}$ is exactly the fluctuating operator given by ${\cal
H}_-={\cal H}=-{\bf I}\frac{d^2}{dz^2} +{\bf V}_{F}(z),$ where
${\bf V}_-={\bf V}_{F}(z)$ is the non-diagonal fluctuation
Hessian. The supersymmetric fluctuation operator partner of ${\cal
H}_-$ is

\begin{equation}
\label{SP} {\cal H}_+={\cal A}^- {\cal A}^+={\cal A}^+ {\cal
A}^-+[{\cal A}^-,{\cal A}^+] ={\cal H}_--{\bf W}^{\prime}(z),
\end{equation}
so that the SUSY partner is given by ${\bf V}_{+}={\bf V}_--{\bf
W}^{\prime}(z).$

Starting with
\begin{equation}
\label{E388} {\cal H}_{-}\Psi^{(n)}_{-} =
E^{(n)}_{-}\Psi^{(n)}_{-} \Longrightarrow  {\cal A}^+ {\cal A}^-
\Psi^{(n)}_{-} = E^{(n)}_{-}\Psi^{(n)}_{-}
\end{equation}
and multiplying (\ref{E388}) from the left by $A^{-}$ we obtain

\begin{equation}
{\cal A}^- {\cal A}^+({\cal A}^- \Psi^{(n)}_{-}) = E^{(n)}_{-}
({\cal A}^- \Psi^{(n)}_{-}) \Rightarrow {\cal H}_{+}({\cal A}^-
\Psi^{(n)}_{-}) = E^{(n)}_{-} (A^- \Psi^{(n)}_{-}). \label{E39}
\end{equation}
Since ${\cal A}^{-} \Psi^{(0)}_{-} = 0,$ comparison of (\ref{E39})
with

\begin{equation}
{\cal H}_{+}\Psi^{(n)}_{+} = {\cal A}^- {\cal A}^+ \Psi^{(n)}_{+}
= E^{(n)}_{+}\Psi^{(n)}_{+}, \label{E40}
\end{equation}
leads to the immediate mapping:

\begin{equation}
E^{(n)}_{+} = E^{(n+1)}_{-}, \quad\Psi^{(n)}_{+}\propto A^{-}
\Psi^{(n+1)}_{-},
 n = 0, 1, 2, \ldots.
\label{E41}
\end{equation}

Repeating the procedure but starting with (\ref{E40}) and
multiplying the same from the left by $A^{+}$ leads to

\begin{equation}
\label{generallabel } {\cal A}^+ {\cal A}^- ({\cal A}^+
\Psi^{(n)}_{+}) = E^{(n)}_{+} ({\cal A}^+ \Psi^{(n)}_{+}),
\label{E42}
\end{equation}
so that it follows from (\ref{E388}), (\ref{E41}) and (\ref{E42})
that

\begin{equation}
\Psi^{(n+1)}_{-}\propto {\cal A}^{+} \Psi^{(n)}_{+} , n = 0, 1, 2,
\ldots. \label{E43}
\end{equation}
The intertwining operator ${\cal A}^{-}({\cal A}^{+})$ converts an
eigenfunction of ${\cal H}_{-}({\cal H}_{+})$ into an
eigenfunction of ${\cal H}_{+}({\cal H}_{-})$ with the same energy
and simultaneously destroys (creates) a node of $\Psi^{(n+1)}_{-}
\left(\Psi^{(n)}_{+} \right).$ These operations just express the
content of the SUSY operations effected by $Q_{+}$ and $Q_{-}$
connecting the bosonic and fermionic sectors of the SUSY
fluctuation operator (\ref{ofs}).

The SUSY analysis presented above in fact enables the generation
of a hierarchy of Hamiltonians with the eigenvalues and the
eigenfunctions of the different members of the hierarchy in a
simple manner. Calling  ${\cal H}_{-}$ as ${\cal H}_{1}$ and
${\cal H}_{+}$ as ${\cal H}_{2}$, and suitably changing the
subscript qualifications, and by repetition of the above procedure
leads to the generation of a hierarchy of Hamiltonians given by

\begin{equation}
{\cal H}_n = -\frac{1}{2}\frac{d^{2}}{dx^{2}} + V_n (x) ={\cal
A}^{+}_n {\cal A}^{-}_n + E^{(0)}_n = {\cal A}^{-}_{n-1} {\cal
A}^{+}_{n-1} + E^{(0)}_{n-1} , \label{E52}
\end{equation}

\begin{equation}
\label{E53} {\cal A}^{\pm}_n =\pm\hbox{{\bf I}}
\frac{d}{dz}+\hbox{{\bf W}}_n(z), \quad \quad \hbox{{\bf
W}}_n^{\dagger}(z)=\hbox{{\bf W}}_n(z)
\end{equation}
whose spectra satisfy the conditions

\begin{equation}
E^{n-1}_{1} = E^{n-2}_{2} = \ldots = E^{(0)}_{n}, \qquad
n=2,3,\ldots, M, \label{E55a}
\end{equation}

\begin{equation}
\Psi^{n-1}_{1}  \propto  {\cal A}^{+}_{1} {\cal A}^{+}_{2} \ldots
{\cal A}^{+}_{n-1} \Psi^{(0)}_n.
\label{E55b}
\end{equation}
Note that the nth-member of the hierarchy has the same eigenvalue
spectrum as the first member ${\cal H}_{1}$ except for the missing
of the first $(n-1)$ eigenvalues of ${\cal H}_{1}$. The
(n-1)th-excited state of ${\cal H}_{1}$ is degenerate with the
ground state of ${\cal H}_{n}$ and can be constructed with the use
of (\ref{E55b}) that involves the knowledge of $A_{i}(i=1,2,
\ldots, n-1)$ and $\Psi^{(0)}_{n}.$

\section{A potential model of three scalar fields}

Let us consider the following generalized potential in terms
of bosonic fields only

\begin{eqnarray}
\label{EP3}
V=&&V(\phi_1,\phi_2,\phi_3)= \frac
12\left(\lambda\phi_1^{2}+\frac{\alpha}{2}\phi_2^2
+\frac{\alpha}{2}\phi_3^2 -\frac{m^2}{\lambda}\right)^2\nonumber\\
+&&\frac 12 \left(-\alpha\phi_1\phi_2+ \beta_2\phi_3^2 -
\beta_2\right)^2,
\nonumber\\
+&&\frac
12\phi_3^2\left(\alpha\phi_1-2\beta_2\phi_2-\alpha\beta_1\right)^2
\end{eqnarray}
where $\alpha>0$ and $\beta_i\geq 0.$  Note that the symmetry
$Z_2$x$Z_2$ is only preserved if $\phi_2=0$ or $\beta_1=\beta_2=0$
condition is satisfied. When $\phi_3=0$ this potential becomes the
two-field potential model recently investigated
\cite{guila02,ger06}.

From (\ref{EMG}) and (\ref{EP3}) the equations of motion under
static limit, for each bosonic component-field, become

\begin{eqnarray}
\label{EM1}
\phi_1^{\prime\prime}=&&2\lambda\phi_1\left(\lambda\phi_1^{2}
+\frac{\alpha}{2}\phi_2^2
+\frac{\alpha}{2}\phi_3^2 -\frac{m^2}{\lambda}\right)\nonumber\\
-&&\alpha\phi_2\left(\beta_2\phi_3^2
-\alpha\phi_2\phi_1-\beta_2\right)\nonumber\\
-&&\alpha\phi_3^ 2
\left(2\beta_2\phi_2-\alpha\phi_1+\alpha\beta_1\right),
\end{eqnarray}

\begin{eqnarray}
\label{EM2}
\phi_2^{\prime\prime}=&&\alpha\phi_2\left(\lambda\phi_1^{2}+
\frac{\alpha}{2}\phi_2^2
+\frac{\alpha}{2}\phi_3^2 -\frac{m^2}{\lambda}\right)\nonumber\\
-&&\alpha\phi_1
\left(\beta_2\phi_3^2-\alpha\phi_1\phi_2-\beta_2\right)\nonumber\\
-&&2\beta_2\phi_3^2
\left(-2\beta_2\phi_2+\alpha\phi_1-\alpha\beta_1\right)
\end{eqnarray}

\begin{eqnarray}
\label{EM3}
\phi_3^{\prime\prime}=&&\alpha\phi_3\left(\lambda\phi_1^{2}+
\frac{\alpha}{2}\phi_2^2
+\frac{\alpha}{2}\phi_3^2 -\frac{m^2}{\lambda}\right)\nonumber\\
+&&2\beta_2\phi_3\left(\beta_2\phi_3^2-\alpha\phi_1\phi_2-
\beta_2\right)\nonumber\\
+&&\phi_3
\left(-2\beta_2\phi_2+\alpha\phi_1-\alpha\beta_1\right)^2.
\end{eqnarray}

The corresponding superpotential model in field theory
is given by

\begin{equation}
\label{SM3}
W(\phi_j)=\frac{m^2}{\lambda}\phi_1-
\frac{\lambda}{3}\phi_1^3-\frac{\alpha}{2}\phi_1\phi_2^2
-\frac{\alpha}{2}\phi_1\phi_3^2+\beta_2\phi_2\phi_3^2-\beta_2\phi_2+ \frac
12\alpha\beta_1\phi_3^2,
\end{equation}
where $\alpha>0$ and $\beta_i>0.$ Thus, it is required that all $\phi_j,
\quad j=1,2,3$ satisfy the BPS state conditions:

\begin{eqnarray}
\label{EBPS3}
  \phi_1^{\prime}&&=-\left(\lambda\phi_1^{2}+\frac{\alpha}{2}\phi_2^2
+\frac{\alpha}{2}\phi_3^2 -\frac{m^2}{\lambda}\right)  \nonumber\\
\phi_2^{\prime}&&=\beta_2\phi_3^2-\alpha\phi_2\phi_1-\beta_2 \nonumber\\
\phi_3^{\prime}&&=-\phi_3(\alpha\phi_1-2\beta_2\phi_2-\alpha\beta_1).
\end{eqnarray}

Note that the BPS states saturate the lower bound so that
$E_{BPS}=|W_{ij}|$ is the central charge of the realization of $N=1$
SUSY in 1+1 dimensions. Thus, the vacua are determined by the
extrema of the superpotential

\begin{equation}
\label{cv}
  \frac{\partial W}{\partial\phi_j}=0, \quad j=1, 2, 3
\end{equation}
which provides the possible vacuum states.

Let us now consider a projection on the $(\phi_1, \phi_2)$ plane
in order to find an explicit form of domain walls. In this case,
if we choose $\phi_3=0$ and $\beta_2=0$, the superpotential $W(\phi_i)$, becomes

\begin{equation}
 W(\phi_i)=\frac{m^2}{\lambda}\phi_1-
\frac{\lambda}{3}\phi_1^3- \frac{\alpha}{2} \phi_1 \phi_2^2,
\end{equation}
which has been discussed recently and some orbits for this
projection have been investigated in references
\cite{guila02,ger06}. Indeed, using the vacuum states, the
following trial orbit,

\begin{equation}
G(\phi_1, \phi_2)=c_1\phi_1^2+c_2\phi_2^2+c_3=0,
\end{equation}
from $\frac{dG}{dx}=\frac{\partial
G}{\partial\phi_1}\phi_1^{\prime}+\frac{\partial
G}{\partial\phi_2}\phi_2^{\prime}=0,
\phi_i^{\prime}=\frac{d\phi_i}{dz}$ and the BPS states
(\ref{EBPS3}), we obtain the constant coefficients, $c_i.$
Therefore, the ground state for this projection on the $(\phi_1,
\phi_2)$ plane becomes

\begin{equation}
\label{gsp}
\Psi_-^{(0)}(z)=\left(
\begin{array}{ccc}
-\lambda\phi_1^{2}(z)-\frac{\alpha}{2}\phi_2^2(z) +\frac{m^2}{\lambda} \\
-\alpha\phi_2(z)\phi_1(z) \\
0
\end{array}\right)
\end{equation}
which represents the bosonic three-component zero mode.
 However,
we have seen that if $\Psi^{(0)}_-$ is a normalizable
three-component eigenstate, one cannot write $\Psi^{(0)}_+$ in
terms of $\Psi^{(0)}_-$ in a similar manner to ordinary
supersymmetric quantum mechanics. Also, $A^{-} =
\psi^{(0)}_-\left(-\frac{d}{dx}\right) \frac{1}{\psi^{(0)}_{-}},$
is only valid for the unidimensional case with one-component
eigenstate. There the superpotential is given by
\begin{equation}
\label{spo}
W(x)=\frac{d}{dx}\ln \psi_-^{(0)}(x).
\end{equation}
 This can not be seen in the example treated here of the
classical stability analysis for three coupled real scalar fields.

The matrix superpotentials in SUSY QM for other projections of the
system considered here will be investigated in a forthcoming
paper, which can not be written as Eq. (\ref{spo}).

\section{Conclusion}

\paragraph*{}

In this paper, we consider the classical stability analysis for
BPS domain walls associated with a potential model of three
coupled real scalar fields, which holds for non-ordinary
supersymmetry (SUSY). The approach of effective quantum mechanics
provides a realization of the SUSY algebra in the three-domain wall
sector of the non-relativistic formalism.

The tensions can be deduced from the charge central properties in a
model that present $N=1$ SUSY, which depend on the manifold of
vacuum states, $T_{ij}=|W[M_j]-W[M_i]|,$ where $M_i$ and $M_j$
represent the vacuum states.

We have shown that the positive potentials with a square form lead
to three-component non-negative normal modes $\omega_n{^2} \geq 0$,
analogous to the case with a single field \cite{R95}, so that the
linear stability of the Schr\"odinger-like equations is ensured.

A general three-component zero-mode eigenfunction is deduced.
 The Ricatti equation (\ref{ER})
holds the BPS states only. A detailed analysis in SUSY QM for such
of a potential model will be reported separately.

In a three-field potential model there is a static classical
configuration inside a topological soliton. We point out that the
superpotential model investigated here can be applied in order to
implement new string junctions \cite{jun} by extendeding BPS domain
walls and string theory of three bosonic moduli \cite{string}.
The set of potential BPS junctions that have been identified in
\cite{dani00} contain the junctions that appear in \cite{safin99}.
Also, the BPS saturated objects with axial symmetry (wall
junctions, vortices), in generalized Wess-Zumino models, have been
investigated in Ref. \cite{ShifV}.

\centerline{\bf Acknowledgments}

RLR would like to acknowledge S. Alves for hospitality at CCP-CBPF
of Rio de Janeiro-RJ, Brazil, were the part of this work was
carried out and to J. A. Hela\"yel-Neto and I. V. Vancea for many
stimulating discussions.  This work was partially supported by
Conselho Nacional de desenvolvimento Cient\'\i fico e
Tecnol\'ogico(CNPq) and Funda\c c\~ao de Apoio \`a Pesquisa do Estado da
Para\'\i ba(FAPESQ)/CNPq(PRONEX).

\end{document}